\documentclass[12pt,14paper]{article}
\usepackage[english]{babel}
\usepackage[latin1]{inputenc}
\usepackage[normalem]{ulem}
\usepackage[intlimits]{amsmath}
\usepackage{amssymb}
\usepackage{graphics}
\usepackage{geometry}
\usepackage{color}
\usepackage{verbatim} 
\usepackage{axodraw}
\usepackage{multirow} 
\usepackage{enumitem}

\usepackage{array}
\usepackage{float}
\usepackage{lscape,graphicx} 
\usepackage{amssymb}

\newcommand{\GeV}{\,\mbox{GeV}}

\newcommand{\bea}{\begin{equation}\begin{array}{c}}
\newcommand{\eea}{\end{array}\end{equation}}
\newcommand{\ea}{\end{array}}

\newcommand{\beq}{\begin{equation}}
\newcommand{\eeq}{\end{equation}}
\newcommand{\bad}{\begin{array}{ccc}}

\newcommand{\ba}{\begin{array}{c}}

\begin{document}

\title{\vspace{-2cm} 
{\normalsize
\flushright FLAVOUR(267104)-ERC-61\\
\vspace{-0.4cm} 
\flushright TUM-HEP 918/13\\}
\vspace{0.6cm} 
\bf Cosmological and astrophysical signatures \\ of dark matter annihilations into \\ pseudo-Goldstone bosons\\[8mm]}
\author{Camilo~Garcia-Cely, Alejandro~Ibarra and Emiliano~Molinaro\\[2mm]
{\normalsize\it Physik-Department T30d, Technische Universit\"at M\"unchen,}\\[-0.05cm]
{\it\normalsize James-Franck-Stra\ss{}e, 85748 Garching, Germany}
}

\maketitle

\begin{abstract}
We investigate a model where the dark matter particle is a chiral fermion field charged under a global $U(1)$ symmetry which is assumed to be spontaneously broken, leading to a pseudo-Goldstone boson (PGB). We argue that the dark matter annihilation into PGBs determine the dark matter relic abundance. Besides, we also note that experimental searches for PGBs allow either for a very long lived PGB, with a lifetime much longer than the age of the Universe, or a relatively short lived PGB, with a lifetime shorter than one minute. Hence, two different scenarios arise, producing very different signatures. In the long lived PGB scenario, the PGB might contribute significantly to the radiation energy density of the Universe. On the other hand, in the short lived PGB scenario, and since the decay length is shorter than one parsec, the s-wave annihilation into a PGB and a $CP$ even dark scalar in the Galactic center might lead to an intense box feature in the gamma-ray energy spectrum, provided the PGB decay branching ratio into two photons is sizable. We also analyze the constraints on these two scenarios from thermal production, the Higgs invisible decay width and direct dark matter searches.
\end{abstract}

\section{Introduction}

While there is a strong empirical evidence for the existence of a dark sector beyond the Standard Model (SM) of Particle Physics, very little is known about its matter content or its interactions. It is well established, however, that the dark sector has a matter density today which amounts to approximately the 27\% of the critical density~\cite{Planck2013} and is constituted by at least one very long lived particle (for reviews see \cite{Bertone:2004pz,Bergstrom:2000pn}). 

A plausible scenario for the dark sector consists in postulating a global $U(1)$ symmetry which is spontaneously broken by a scalar field with charge 2 under that symmetry, hence leading to a remnant discrete $Z_2$ symmetry in the Lagrangian. All the fields with even (odd) charge under the global group will acquire, after the spontaneous symmetry breaking, an even (odd) discrete charge under the $Z_2$ transformation~\cite{Krauss:1988zc}. Therefore, the lightest particle with odd charge is absolutely stable and a potential candidate for dark matter. A dark sector with this characteristics was proposed by Weinberg in \cite{Weinberg:2013kea} and has a matter content consisting in a complex scalar field and a Dirac fermion.  Other models following a similar rationale were presented in \cite{Lindner:2011it,JosseMichaux:2011ba}, where the global symmetry was identified with $U(1)_{B-L}$, generating also neutrino masses and the correct baryon asymmetry via leptogenesis, and in \cite{Lee:2012bq,Ibarra:2013eda,Dasgupta:2013cwa}, where the global symmetry was identified with the Peccei-Quinn symmetry.
The model contains a massless Goldstone boson arising from the spontaneous breaking of the global continuous symmetry which, as argued  in \cite{Weinberg:2013kea}, could contribute to the radiation energy density of the Universe. This model was studied in detail in \cite{Garcia-Cely:2013nin}, where it was shown that the Goldstone boson plays a crucial role in the dark matter production. Furthermore, it was shown that this model might lead to observable signatures in direct dark matter searches. In particular, large regions of the parameter space will be probed by the ongoing LUX experiment \cite{Akerib:2013tjd,Akerib:2012ys} or the projected XENON1T \cite{Aprile:2012zx}, under the assumption that the observed dark matter abundance was thermally produced and that the Goldstone boson of the model accounts for the hints for dark radiation reported in \cite{Planck2013}. On the other hand, observing signatures of this model in indirect dark matter search experiments is challenging since all dark matter annihilation processes are p-wave suppressed. Phenomenological analyses of this model have also been presented in \cite{Cheung:2013oya,Anchordoqui:2013pta,Anchordoqui:2013bfa}.

We propose in this paper a variant of this model, where the Dirac fermion is replaced by a chiral fermion. We will show that, due to the explicit $C$ and $P$ breaking, dark matter particles can annihilate in the s-wave into a Goldstone boson and a $CP$ even hidden sector scalar, thus requiring a smaller coupling to reproduce the correct relic abundance than in the Dirac case discussed in \cite{Weinberg:2013kea}. Furthermore, we will consider the situation where the dark Abelian symmetry is not exact (while preserving the exact $Z_2$ symmetry), hence the Goldstone boson is replaced by a (massive) pseudo-Goldstone boson, which can decay into two photons. While the mass and lifetime of a pseudo-Goldstone boson is severely constrained by various observations, two windows remain at the moment viable: one with a lifetime longer than $\sim 10^{20}$ years and one with a lifetime shorter than one minute (see \cite{Hewett:2012ns} for a recent compilation of searches for pseudo-Goldstone bosons). In the former case, the pseudo-Goldstone boson becomes a firm candidate for dark radiation along the lines of \cite{Weinberg:2013kea}, while in the latter case, the decay length is shorter than $\sim 1$ pc and hence the pseudo-Goldstone bosons produced in dark matter annihilations in the Galactic center decay in flight before reaching the Earth, thus producing a gamma-ray flux displaying a characteristic box-shape spectrum and with an intensity that could be at the reach of gamma-ray telescopes. 

The paper is organized as follows. In Section \ref{sec:model} we present our model of the dark sector and in Section \ref{sec:constraints} the constraints on the model parameters from direct dark matter search experiments and the invisible Higgs decay width. In Section \ref{sec:production} we discuss the thermal production of dark matter and we show the existence of a s-wave annihilation channel into Goldstone bosons. 
In Section \ref{sec:PGB} we analyze the two allowed windows in the pseudo-Goldstone boson parameter space and we discuss the cosmological and astrophysical signatures arising in each of the two windows. Lastly, in Section \ref{sec:conclusions} we present our conclusions.

\section{Description of the Dark Sector}
\label{sec:model}

We extend the Standard Model (SM) Lagrangian with one complex scalar field $\phi$, and one chiral fermion field, which we assume for concreteness left-handed, $\psi_{L}$ (the analysis for a right-handed field is analogous). These new fields are SM singlets and are charged under a global $U(1)_\text{DM}$ symmetry, namely  $U(1)_\text{DM}(\psi_{L})=1$ and $U(1)_\text{DM}(\phi)=2$. On the other hand, all the SM fields transform trivially under the additional global symmetry, which could be exact or nearly exact. Let us discuss each case separately.

\subsection{Exact $U(1)_\text{DM}$ symmetry}
\label{subsec:exact}

If the global symmetry is exact, the interaction Lagrangian is
\begin{eqnarray}
	\mathcal{L}  &=&  \mu_{H}^{2}\,H^{\dagger}\,H\,-\,\lambda_{H}\,\left(H^{\dagger}\,H\right)^{2}\,+\,
	\mu_{\phi}^{2}\,\phi^{\dagger}\,\phi\,-\,\lambda_{\phi}\,\left(\phi^{\dagger}\,\phi\right)^{2}\,-\,
	\kappa\,\left(H^{\dagger}\,H\right)\,\left(\phi^{\dagger}\,\phi\right)\nonumber\\
	&& +\,i \overline{\psi_L}\gamma^\mu \partial_\mu \psi_L - \left(\dfrac{f}{\sqrt{2}}\phi \,\overline{\psi_L} \,\psi_L^c + h.c.\right)\,,
\label{lint}
\end{eqnarray}
where $H$ is the SM Higgs doublet. Notice that the complex phase of the coupling constant $f$ can be absorbed by redefining the scalar field $\phi$. As a result, $CP$ is conserved in this model, while $C$ and $P$ are explicitly broken.
Both the scalar field $\phi$ and the neutral component of the Higgs doublet acquire non-zero vacuum expectation values, which spontaneously break the symmetry group $SU(2)_{\rm W}\times U(1)_{\rm Y}\times [U(1)_\text{DM}] \to U(1)_{\rm em} \times Z_{2} $. In order to analyze the physical mass spectrum of the theory, we conveniently parametrize the scalar fields in Eq.~(\ref{lint}) as:
\begin{equation}
H  =\begin{pmatrix} G^+ \\ \frac{v_H+\tilde{h}+i G^0 }{\sqrt{2}} \end{pmatrix} \;, \hspace{40pt} \phi  = \frac{v_\phi+\tilde{\rho}+i\eta}{\sqrt{2}} \;,
\label{fieldHphi}
\end{equation}
where $v_{H}\simeq 246$ GeV. The scalar mass spectrum  consists of a $CP$ odd massless scalar $\eta$, which is the Goldstone boson that arises from the spontaneous breaking of the global $U(1)_\text{DM}$ symmetry, and two $CP$ even massive real scalars, denoted by $h$ and $\rho$ and with mass $m_h$ and $m_\rho$ respectively, which arise from the mixing of the interaction fields $\tilde{h}$ and $\tilde{\rho}$ by means of an angle $\theta$ \cite{Garcia-Cely:2013nin}. The quartic couplings in the Lagrangian Eq.~(\ref{lint}) can then be related to the masses and the mixing angle in the scalar sector by: 
\begin{eqnarray}\label{kappaeq}
\lambda_H  = \frac{m_h^2 \cos^2\theta + m_\rho^2 \sin^2\theta}{2 v_H^2},\hspace{30pt}
\lambda_\phi  = \frac{m_h^2 \sin^2\theta + m_\rho^2 \cos^2\theta}{2 v_\phi^2},&&\nonumber\\
\kappa =\frac{(m_{\rho}^{2}-m_{h}^{2})\,\sin2\theta}{2\,v_{H}\,v_{\phi}}\,.\hspace{100pt} &&
\label{quarticc}
\end{eqnarray}

While the scalar potential of this model is identical to the one considered in \cite{Weinberg:2013kea, Garcia-Cely:2013nin}, the fermionic sector contains significant differences. Indeed, in this model only one Majorana fermion, which we denote by $\chi$, arises after the symmetry breaking. The corresponding mass-eigenstate and Majorana mass are
\begin{equation}
\chi = \psi_L + (\psi_L)^c\,,\quad\quad\quad\quad\quad M_\chi = f v_\phi \,.
\label{masschi}
\end{equation}
With these definitions, the part of the Lagrangian involving $\chi$ can be cast as 
\begin{eqnarray}
{\cal L}_{\chi} &=& \dfrac{i}{2} \overline{\chi}\gamma^\mu \partial_\mu \chi - \dfrac{f}{\sqrt{2}}(\phi \overline{\chi} P_R \chi + \phi^* \overline{\chi} P_L \chi )\,,\label{lintchiralsym}
\end{eqnarray}
which after electroweak symmetry breaking becomes
\begin{eqnarray}
{\cal L}_{\chi}&=& \dfrac{1}{2} ( i \overline{\chi}\gamma^\mu \partial_\mu \chi - M_\chi \overline{\chi} \chi ) - \dfrac{f}{2}( (-\sin\theta h + \cos\theta \rho) \overline{\chi}\chi +  i \eta \overline{\chi}\gamma^5 \chi  )\,.
\label{lintchiral}
\end{eqnarray}
From Eqs.~(\ref{quarticc}) and (\ref{masschi}), it follows that there are four unknown independent parameters describing the dark sector, which can be taken as $m_\rho$, $\theta$, $M_\chi$ and $f$. 

Notice that the Lagrangian in Eq.~(\ref{lintchiralsym}) is invariant under $U(1)_\text{DM}$ upon the field transformations $\psi_L \to e^{i\alpha} \psi_L$, or equivalently,  $\chi \to e^{-i\alpha\gamma^5} \chi$.~\footnote{This transformation also leaves the Majorana condition $\chi= \chi^{c}$ invariant.} On the other hand, after the symmetry breaking, and due to the presence of the Majorana mass $M_{\chi}$, the Lagrangian is no longer invariant under the continuous transformation although, as expected, it preserves  a remnant discrete symmetry $\chi\to -\chi$. The Majorana field $\chi$ then describes a stable neutral particle and is therefore a viable dark matter candidate.

We have assumed here the simplest scenario where the Majorana field $\chi$ transforms as a singlet of the global symmetry. More complicated scenarios can be constructed with identical properties regarding the dark matter stability, for example by assuming that the Majorana field transforms  as a doublet of a global symmetry $SO(2)\cong U(1)$. This scenario is equivalent to the axion-mediated dark matter model discussed in \cite{Lee:2012bq,Ibarra:2013eda}, in which the two components of the doublet form a  Dirac fermion.

\subsection{Nearly exact $U(1)_\text{DM}$ symmetry}
\label{subsec:nearly-exact}

We consider now the situation in which the global $U(1)_\text{DM}$ is not an exact symmetry of the Lagrangian. However, we assume that the Lagrangian Eq.~(\ref{lint}) still describes to a very good approximation the phenomenology of the dark sector, $i.e.$, that the $U(1)_\text{DM}$ is a nearly exact symmetry. In particular, we demand that the stability of the dark matter is not affected by the explicit  breaking of the global symmetry, that is we  postulate that $Z_{2}$ is a symmetry of the part of the Lagrangian that breaks the global $U(1)_\text{DM}$ symmetry explicitly. If this is the case, the $\eta$ particle is a pseudo-Goldstone boson with a mass $m_\eta$ much smaller than the scale at which the global symmetry spontaneously breaks, namely $m_\eta \ll v_\phi$. We can therefore reasonably assume that $m_\eta \ll m_\rho$, and neglect the pseudo-Goldstone mass henceforth. 

 An important difference of this scenario compared to the one described in Subsection \ref{subsec:exact} is that, when the symmetry is nearly exact, the (massive) pseudo-Goldstone boson might decay into two photons. Such process is induced by the effective operator
\begin{equation}
\mathcal{L}_{eff} \supseteq -\frac{1}{4} \, g_{\eta\gamma} \, \epsilon^{\mu\nu\alpha\beta}\, F_{\mu\nu}\,F_{\alpha\beta}\, \eta \;,
\label{Leff}
\end{equation}
where $g_{\eta\gamma}$ is a coupling constant with dimensions of inverse of energy and $F_{\mu\nu}$ is the electromagnetic field strength tensor. This Lagrangian arises in dark sectors with new chiral fermion representations charged under the SM group with masses of order $\Lambda \gg v_\phi$, that make the global $U(1)_\text{DM}$ symmetry anomalous. Consequently, in analogy to the neutral pions in the Standard Model, an effective coupling between the pseudo-Goldstone boson $\eta$ and the gauge fields might be generated by non-perturbative processes involving the new heavy degrees of freedom.
For instance, this happens in axion-mediated dark matter models where the pseudo-Goldstone boson, the axion, arises from the spontaneous breaking of an anomalous Peccei-Quinn symmetry (see, $e.g.$, \cite{Lee:2012bq,Ibarra:2013eda}). In this paper we adopt a phenomenological approach and simply assume that the operator given in Eq.~(\ref{Leff}) exists, without specifying the new physics responsible for its origin.

\section{Constraints from Direct Searches and the Invisible Higgs Decay Width}
\label{sec:constraints}

The scalar $\rho$ and the Higgs boson $h$ might decay into two dark matter particles, two \mbox{(pseudo-)Goldstone} bosons or SM particles. The relevant decay widths for $\rho$ read
\begin{eqnarray}
	&&\Gamma(\rho\to \eta\,\eta)  \;= \; \frac{f^{2}\,r^2}{32\,\pi }\,m_{\rho}\, \cos^{2}\theta,\label{r0invdec}\\
	&&\Gamma\left(\rho\to \chi\chi\right) \; = \;\frac{f^{2}}{16\,\pi }\, \left(1-\frac{4}{r^2}\right)^{3/2}\,m_\rho\,\cos^{2}\theta\,,\label{rtoDMDM}\\
	&&\Gamma\left(\rho\to \text{SM~particles}\right) \; = \; \sin^{2}\theta\,\Gamma^{\text{SM}}\left(\text{Higgs}\to \text{SM\;particles}\right)\,,
\end{eqnarray}
where $r \equiv m_\rho/M_\chi$. The corresponding expressions for $h$ are obtained by exchanging $\cos\theta$ for $\sin\theta$ and $m_\rho$ for $m_h$. In addition, the heaviest $CP$ even scalar can decay into the
lightest one with a phase space suppressed rate (see \cite{Garcia-Cely:2013nin} for details). From these equations and the experimental upper limit on the invisible decay width of the Higgs boson (see, $e.g.$, \cite{Belanger:2013kya}) , it follows that the mixing angle $\theta$ is bounded from above by \cite{Weinberg:2013kea}:
\begin{equation}
	\left|\tan\theta\right|	\;\lesssim\;2.2\times 10^{-3}\,\left(\frac{v_{\phi}}{10\,{\rm GeV}}\right)\,\quad\text{or }\quad
	f\left|\sin2\theta\right| \;\lesssim\;4.4\times 10^{-3}\,\left(\frac{M_{\chi}}{10\,{\rm GeV}}\right)\,,\label{thetabound}
\end{equation}
where in the last expression we have used Eq.~(\ref{masschi}). 

\begin{figure}[t!]
\begin{center}
\includegraphics[width=12cm]{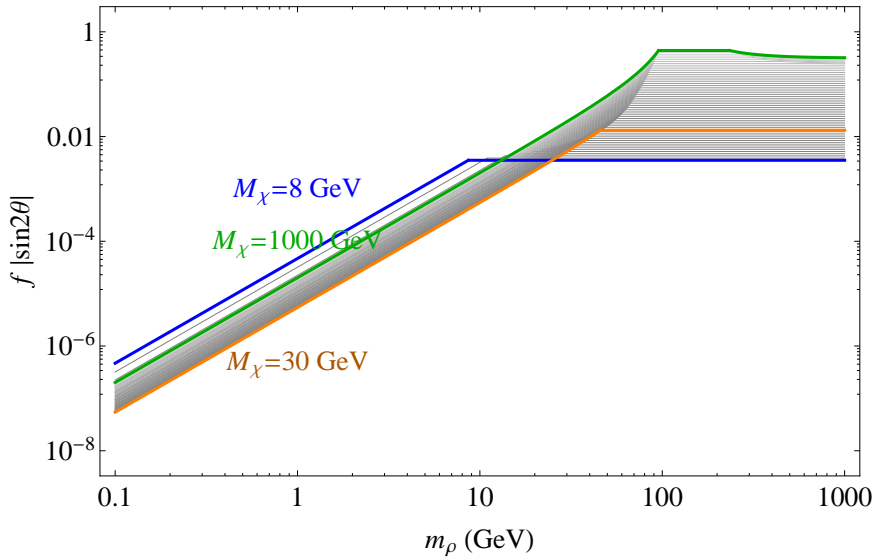}
\caption{Combined upper limit on $f |\sin2\theta|$ from direct dark matter searches and the invisible Higgs decay width as a function of the mass of the $CP$ even dark scalar for various values of the dark matter mass.}
\label{figure:fsin2thetaBound}
\end{center}
\end{figure}

Direct dark matter searches constrain the same combination of parameters, $f\left|\sin2\theta\right|$. The calculation of the scattering cross-section of dark matter off nucleons is analogous as in \cite{Garcia-Cely:2013nin}, the result being
\begin{eqnarray}
\sigma_{\chi\,N}& =& C^2\, \frac{ \, m_N^4 \,M_\chi^2}{4\pi \,v_H^2\, (M_\chi+m_N)^2} \,\left(\frac{1}{m_h^2}-\frac{1}{m_\rho^2}\right)^2 (f\,\sin2\theta)^2\,,
\label{DirectDetectionEq}
\end{eqnarray}
where $m_N$ denotes the nucleon mass and $C\simeq 0.27$ \cite{Belanger:2013oya} is a constant that depends on the nucleon matrix element. In Fig.~\ref{figure:fsin2thetaBound} we show, as black lines, the upper limit on $f|\sin2\theta|$ as a function of $m_\rho$ for various dark matter masses between $8 ~\GeV$ and $1000 ~\GeV$ from the invisible Higgs decay width, Eq.~(\ref{thetabound}), and from the LUX experiment ~\cite{Akerib:2013tjd}, Eq.~(\ref{DirectDetectionEq}); the blue, orange and green lines correspond to $M_\chi=8, 30$ and $1000~\GeV$ respectively. It follows from the plot that for $\rho$ masses below $10$ GeV the bound on $f|\sin2\theta|$ is determined by direct detection experiments, whereas for $m_\rho$ larger than $100$ GeV, by the upper limit on the invisible Higgs decay width (dominated in this mass range by $h\rightarrow \eta\eta$).

\section{Thermal Production of Dark Matter}
\label{sec:production}

The thermal production of dark matter is expected to be dominated by annihilation channels involving the $\rho$ scalar and the (pseudo-)Goldstone boson, similarly to \cite{Garcia-Cely:2013nin}; the relevant diagrams are shown in Fig.~\ref{figure:diagrams} and the expressions for the corresponding cross-sections are reported in the Appendix. On the other hand, annihilations into SM particles are expected to have a fairly suppressed rate,  due to the smallness of the mixing angle $\theta$, except when the dark matter annihilation proceeds via resonant s-channel exchange of $CP$ even scalars, that is, either $\rho$ or $h$.

\begin{figure}[t!]
\begin{center}
% diagrams for process \chi,\chi -> \rho,\eta                            
%\documentstyle[axodraw]{article}
%\begin{document}
{
\unitlength=1.0 pt
\SetScale{1.0}
\SetWidth{0.7}      % line    size control
\scriptsize    %  letter  size control
\qquad\allowbreak
%  diagram # 1
\begin{picture}(96,38)(0,0)
\Line(48.0,35.0)(24.0,35.0) 
\Text(24.0,35.0)[r]{$\chi$}
\Line(48.0,35.0)(48.0,11.0) 
\Text(49.0,24.0)[l]{$\chi$}
\DashLine(48.0,35.0)(72.0,35.0){3.0}
\Text(72.0,35.0)[l]{$\rho$}
\Line(48.0,11.0)(24.0,11.0) 
\Text(24.0,11.0)[r]{$\chi$}
\DashLine(48.0,11.0)(72.0,11.0){3.0}
\Text(72.0,11.0)[l]{$\eta$}
\end{picture} 
\qquad\allowbreak
%  diagram # 2
\begin{picture}(96,38)(0,0)
\Line(48.0,35.0)(24.0,35.0) 
\Text(24.0,35.0)[r]{$\chi$}
\Line(48.0,35.0)(48.0,11.0) 
\Text(49.0,24.0)[l]{$\chi$}
\DashLine(48.0,35.0)(72.0,35.0){3.0}
\Text(72.0,35.0)[l]{$\eta$}
\Line(48.0,11.0)(24.0,11.0) 
\Text(24.0,11.0)[r]{$\chi$}
\DashLine(48.0,11.0)(72.0,11.0){3.0}
\Text(72.0,11.0)[l]{$\rho$}
\end{picture}
\qquad\allowbreak
%  diagram # 3
\begin{picture}(96,38)(0,0)
\Line(36.0,23.0)(12.0,35.0) 
\Text(12.0,35.0)[r]{$\chi$}
\Line(36.0,23.0)(12.0,11.0) 
\Text(12.0,11.0)[r]{$\chi$}
\DashLine(36.0,23.0)(60.0,23.0){3.0}
\Text(49.0,24.0)[b]{$\eta$}
\DashLine(60.0,23.0)(84.0,35.0){3.0}
\Text(84.0,35.0)[l]{$\eta$}
\DashLine(60.0,23.0)(84.0,11.0){3.0}
\Text(84.0,11.0)[l]{$\rho$}
\end{picture} 
}\\
% diagrams for process \chi,\chi -> \rho,\rho                            
%\documentstyle[axodraw]{article}
%\begin{document}
{
\unitlength=1.0 pt
\SetScale{1.0}
\SetWidth{0.7}      % line    size control
\scriptsize    %  letter  size control
\begin{comment}
{} \qquad\allowbreak
%  diagram # 1
\begin{picture}(96,38)(0,0)
\Line(36.0,23.0)(12.0,35.0) 
\Text(12.0,35.0)[r]{$\chi$}
\Line(36.0,23.0)(12.0,11.0) 
\Text(12.0,11.0)[r]{$\chi$}
\DashLine(36.0,23.0)(60.0,23.0){3.0}
\Text(49.0,24.0)[b]{$h$}
\DashLine(60.0,23.0)(84.0,35.0){3.0}
\Text(84.0,35.0)[l]{$\rho$}
\DashLine(60.0,23.0)(84.0,11.0){3.0}
\Text(84.0,11.0)[l]{$\rho$}
\end{picture} 
\end{comment}
\qquad\allowbreak
%  diagram # 2
\begin{picture}(96,38)(0,0)
\Line(48.0,35.0)(24.0,35.0) 
\Text(24.0,35.0)[r]{$\chi$}
\Line(48.0,35.0)(48.0,11.0) 
\Text(49.0,24.0)[l]{$\chi$}
\DashLine(48.0,35.0)(72.0,35.0){3.0}
\Text(72.0,35.0)[l]{$\rho$}
\Line(48.0,11.0)(24.0,11.0) 
\Text(24.0,11.0)[r]{$\chi$}
\DashLine(48.0,11.0)(72.0,11.0){3.0}
\Text(72.0,11.0)[l]{$\rho$}
\end{picture} 
\qquad\allowbreak
%  diagram # 3
\begin{picture}(96,38)(0,0)
\Line(36.0,23.0)(12.0,35.0) 
\Text(12.0,35.0)[r]{$\chi$}
\Line(36.0,23.0)(12.0,11.0) 
\Text(12.0,11.0)[r]{$\chi$}
\DashLine(36.0,23.0)(60.0,23.0){3.0}
\Text(49.0,24.0)[b]{$\rho$}
\DashLine(60.0,23.0)(84.0,35.0){3.0}
\Text(84.0,35.0)[l]{$\rho$}
\DashLine(60.0,23.0)(84.0,11.0){3.0}
\Text(84.0,11.0)[l]{$\rho$}
\end{picture} \ 
}\\
%\end{document}
% diagrams for process \chi,\chi -> \eta,\eta                            
%\documentstyle[axodraw]{article}
%\begin{document}
{
\unitlength=1.0 pt
\SetScale{1.0}
\SetWidth{0.7}      % line    size control
\scriptsize    %  letter  size control
\begin{comment}
\qquad\allowbreak
%  diagram # 1
\begin{picture}(96,38)(0,0)
\Line(36.0,23.0)(12.0,35.0) 
\Text(12.0,35.0)[r]{$\chi$}
\Line(36.0,23.0)(12.0,11.0) 
\Text(12.0,11.0)[r]{$\chi$}
\DashLine(36.0,23.0)(60.0,23.0){3.0}
\Text(49.0,24.0)[b]{$h$}
\DashLine(60.0,23.0)(84.0,35.0){3.0}
\Text(84.0,35.0)[l]{$\eta$}
\DashLine(60.0,23.0)(84.0,11.0){3.0}
\Text(84.0,11.0)[l]{$\eta$}
\end{picture} 
\end{comment}
\qquad\allowbreak
%  diagram # 3
\begin{picture}(96,38)(0,0)
\Line(48.0,35.0)(24.0,35.0) 
\Text(24.0,35.0)[r]{$\chi$}
\Line(48.0,35.0)(48.0,11.0) 
\Text(49.0,24.0)[l]{$\chi$}
\DashLine(48.0,35.0)(72.0,35.0){3.0}
\Text(72.0,35.0)[l]{$\eta$}
\Line(48.0,11.0)(24.0,11.0) 
\Text(24.0,11.0)[r]{$\chi$}
\DashLine(48.0,11.0)(72.0,11.0){3.0}
\Text(72.0,11.0)[l]{$\eta$}
\end{picture} 
\qquad\allowbreak
%  diagram # 2
\begin{picture}(96,38)(0,0)
\Line(36.0,23.0)(12.0,35.0) 
\Text(12.0,35.0)[r]{$\chi$}
\Line(36.0,23.0)(12.0,11.0) 
\Text(12.0,11.0)[r]{$\chi$}
\DashLine(36.0,23.0)(60.0,23.0){3.0}
\Text(49.0,24.0)[b]{$\rho$}
\DashLine(60.0,23.0)(84.0,35.0){3.0}
\Text(84.0,35.0)[l]{$\eta$}
\DashLine(60.0,23.0)(84.0,11.0){3.0}
\Text(84.0,11.0)[l]{$\eta$}
\end{picture} 
}
\end{center}
\caption{Relevant diagrams for dark matter production in the limit $\theta \ll 1 $. The process into $\rho\,\eta$ (first row) proceeds via s-wave whereas the other ones are p-wave suppressed.}
\label{figure:diagrams}
\end{figure}
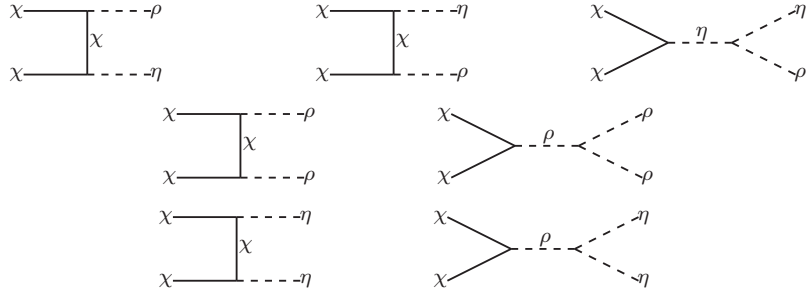

\begin{figure}[t!]
\begin{center}
\includegraphics[width=12cm]{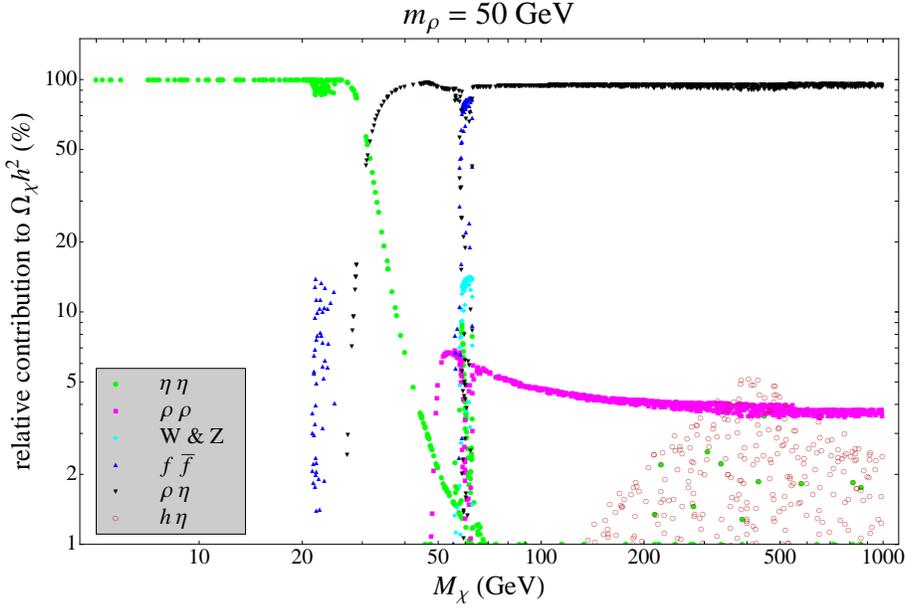}
\caption{Relative contribution to the dark matter relic density from various annihilation channels as a function of the dark matter mass, assuming $m_\rho=50$ GeV.}
\label{figure:scan}
\end{center}
\end{figure}

These expectations are confirmed by our numerical analysis. We have performed scans over the four dimensional parameter space spanned by  $m_{\rho}$, $M_\chi$, $f$ and $\theta$. More specifically, we have performed a logarithmic scan of $m_{\rho}$ between 200 MeV and 1 TeV, $M_\chi$ between 1 GeV and 1 TeV, $f$ between $10^{-2}$ and $4\pi$ and $|\tan\theta|$ between 0 and the maximal value allowed by the invisible decay width of the Higgs, given in Eq.~(\ref{thetabound}) with $v_\phi=M_\chi/f$ (see  Eq.~(\ref{masschi})). We have also checked that the quartic couplings necessary to produce these parameters, inferred from Eq.~(\ref{quarticc}), are smaller than 4$\pi$, in order to ensure perturbativity. We have then calculated for each point the dark matter relic density using micrOMEGAs 3.1~\cite{Belanger:2013oya}, working under an implementation of our model made with FeynRules~\cite{Christensen:2008py}, and we have selected only those points for which $\Omega_{\text{DM}} h^2=0.1199\pm 0.0027$ within $3\sigma$. We report the results of one scan in Fig.~\ref{figure:scan}, where we show the relative contribution to the relic density of each annihilation process for the concrete case $m_{\rho}=50$ GeV. Indeed, the dominant channel is $\chi\chi \to \rho\,\eta$, when this is kinematically open, $i.e.$ for $m_{\rho} <2 M_{\chi}$, while  $\chi\chi \to \eta \eta$ dominates when $m_\rho>2 M_{\chi}$. It is important to note that for certain values of $m_\rho$ threshold effects or resonant effects can have a dramatic impact in the calculation of the relic density, concretely when $m_\rho\approx M_\chi $, close to the threshold of the process $\chi\chi \to \rho \rho$, or when  $m_\rho \approx 2M_\chi $, close to the threshold of  $\chi\chi \to \rho \eta$ and where moreover the process $\chi\chi\rightarrow \eta\eta$ via the s-channel mediation of $\rho$ is resonantly enhanced. Resonance effects are manifest in Fig.~\ref{figure:scan} at $M_{\chi}= m_{h}/2\approx 63$ GeV and $M_{\chi}=m_{\rho}/2\approx 25$ GeV, where the Higgs and $\rho$ resonances take place, respectively.

\begin{figure}[t!]
\begin{center}
\includegraphics[width=12cm]{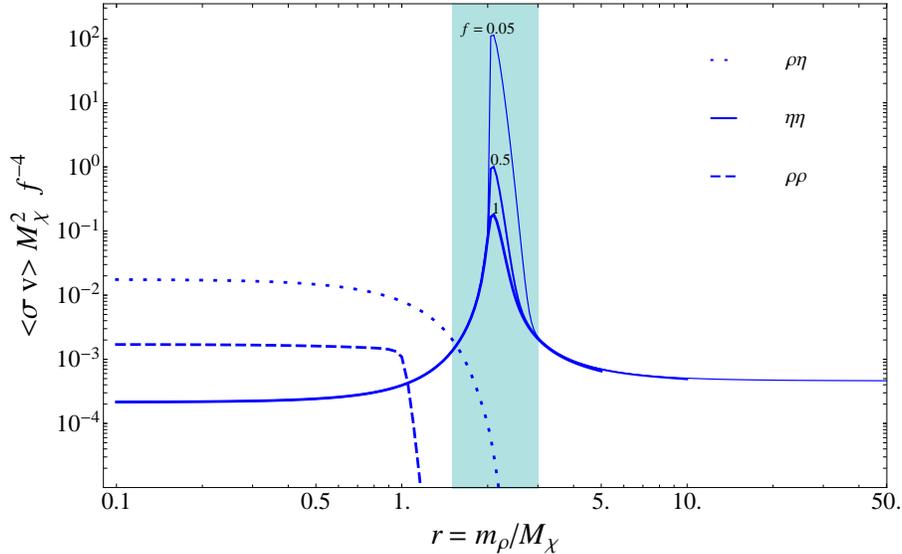}
\caption{Thermally averaged cross-section  $\langle\sigma v\rangle$ for the annihilation channels $\chi\chi\rightarrow \rho \eta,~\eta\eta,~\rho\rho$ as a function of $r\equiv m_\rho/M_\chi$ at the typical freeze-out temperature $T = M_\chi/20$. The resonant behavior of the annihilation into $\eta\eta$ at $r=2$ is due to the s-channel mediation of a $\rho$, with a width determined by the coupling constant $f$.}
\label{figure:thermalsigmav}
\end{center}
\end{figure}

In order to determine the precise regions where threshold and resonance effects have an important impact on the relic density, we have calculated the thermal average of the annihilation cross-sections as a function of $r=m_\rho/M_\chi$; the result is shown in Fig.~\ref{figure:thermalsigmav} for a typical freeze-out temperature, $T \sim M_\chi/20$, and for various values of $f$ which, following Eqs.~(\ref{r0invdec}) and (\ref{rtoDMDM}),  determine the width of $\rho$. As apparent from the plot, the threshold and resonant effects are most relevant in the region $1.5 \lesssim r \lesssim 3$. Furthermore, for $r\lesssim 1.5$ the largest annihilation cross-section corresponds to the process $\chi\chi\rightarrow \rho\eta$, while for  $r\gtrsim 3$ to $\chi\chi\rightarrow \eta\eta$. Notice that, for a given coupling $f$, the upper limit  $r \lesssim  \sqrt{8\pi}/f$ must hold from the requirement of perturbativity, as also reflected in Fig.~\ref{figure:thermalsigmav}.

In the regions where both resonance and threshold effects are negligible, namely $r \lesssim 1.5$ or $r \gtrsim 3$, the relic abundance can be accurately calculated using the instantaneous freeze-out approximation~\cite{Griest:1990kh}. Casting the annihilation cross-section in the form $\sigma v = a+b v^2$, the relic density can be approximated by
\begin{eqnarray}
	\Omega h^{2} & \simeq & \frac{\left(1.07\times 10^{9} \,\text{GeV}^{-1}\right) \,x_f}{g_{*}(x_{f})^{1/2}\,m_{\text{Pl}} \left(a+3b/x_f\right) }\,,\label{omegah2}
\label{Omegah2}
\end{eqnarray}
where typically $x_{f} =M_\chi/T_{f} \approx 20-30$ for WIMP dark matter and $g_{*}(x_{f})$ is the number of relativistic degrees of freedom at the freeze-out temperature. 

In the regime $r\lesssim 1.5$ the three annihilation processes into dark sector particles $\chi\chi\rightarrow \rho\rho,~\eta\eta,~\rho \eta$ and $\chi\chi\to$ SM SM are all kinematically accessible. The annihilations into $\rho\rho$ and $\eta\eta$ are, however, p-wave suppressed and can be safely neglected in the calculation of the relic density. This can be understood analyzing the $CP$ of the initial and final states. We use the standard notation $S$, $L$ and $J$ for the spin, the orbital and the total angular momenta  with a subscript $i$ or $f$ for the corresponding quantities of the initial or final state. Then, the $CP$ eigenvalues of the initial and final states are $(-1)^{L_i+1}$ and $(-1)^{L_f}$. $CP$ conservation thus implies that $|L_f - L_i|$ is an odd number. In addition, since $\rho$ and $\eta$ are scalars, we have $J_i=L_f$. If the s-wave were allowed $L_i=0$ and  $J_i=S_i$. As a result we could only have $S_i=1$ and $L_i=0$, which is impossible for a pair of Majorana fermions due to the Pauli exclusion principle. The only possibility is then $L_i \geq 1$ and hence the cross-sections are p-wave suppressed. Explicitly, they read
\begin{eqnarray}
\sigma v(\chi\chi \to \rho\rho)\;=\;\frac{f^4 v^2 \sqrt{1-r^2} }{384 \pi M_{\chi }^2 }\frac{\left(3 r^4-8 r^2+8\right) \left(9 r^8-64 r^6+200 r^4-352 r^2+288\right)}{\left(r^2-4\right)^2
   \left(r^2-2\right)^4}\,,\label{rhorho}
 \end{eqnarray}
 \begin{eqnarray} 
\sigma v(\chi\chi \to \eta\eta)\;=\;\frac{f^4v^2}{192 \pi  M_{\chi }^2 \left(r^2-4\right)^2}\left(8+r^4\right)\,,
\label{etaeta}
\end{eqnarray}
which are manifestly velocity suppressed. In contrast, for the annihilation into $\rho\eta$ the $CP$ eigenvalues of the initial and final states are $(-1)^{L_i+1}$ and $(-1)^{L_f+1}$. We again have $J_i=L_f$, and therefore $|J_i-L_i|$ is an even number. $CP$ conservation therefore allows the s-wave channel if $J_i$ is even. The corresponding cross-section is
\begin{equation}
\sigma v(\chi\chi \to \rho\eta)\;=\;\frac{f^4}{16 \pi  M_{\chi }^2}\left(1-\frac{r^2}{4}\right)^3\,.
\label{rhoeta}
\end{equation}
Lastly, annihilations into SM particles are p-wave suppressed,  due to $CP$ conservation, and are moreover $\theta$-suppressed. Therefore, they can be safely neglected in our analysis. Hence, in the regime $r\lesssim 1.5$ the relevant process for the calculation of the relic density is the annihilation $\chi\chi \to \rho\eta$.

On the other hand, in the regime $r\gtrsim 3$, the only kinematically open channels are $\chi\chi\rightarrow \eta \eta$ and $\chi\chi\rightarrow {\rm SM}\,{\rm SM}$. Both processes are p-wave suppressed, however the latter has an additional $\theta$-suppression. Therefore, the dominant annihilation process is in this case into $\eta\eta$ with a cross-section given in Eq.~(\ref{etaeta}). 

Using Eq.~(\ref{Omegah2}) it is then possible to estimate the value of the dark matter coupling $f$ as function of $r$ and $M_\chi$ (and $x_f$) leading to the observed dark matter abundance  $\Omega_{\text{DM}} h^2\simeq 0.12$ in the regime $r\lesssim 1.5$ (where the annihilation into $\rho\, \eta$ with a cross-section given by Eq.~(\ref{rhoeta}) determines the dark matter freeze-out) and in the regime $r\gtrsim 3$ (where the annihilation into $\eta\eta$ is the relevant one, with cross-section given by Eq.~(\ref{etaeta})). The coupling reads:
\begin{equation}
f \simeq \left\{ \begin{array}{c}
 0.39 \left(\frac{x_f}{g_{*}(x_{f})^{1/2}  \,\left(4-r^2\right)^3  }\right)^{1/4} \left(\frac{M_\chi}{100~\text{GeV}}\right)^{1/2}  \text{\quad,\quad if\quad}r \lesssim 1.5\\
0.20 \left(\frac{\left(4-r^2\right)^2 x_f^2}{g_{*}(x_{f})^{1/2} \,\left(8+r^4\right)   }\right)^{1/4} \left(\frac{M_\chi}{100~\text{GeV}}\right)^{1/2} \text{\quad,\quad if\quad}r \gtrsim 3
\end{array} \right. \,.
\label{expf}
\end{equation}
\begin{figure}[t]
\begin{center}
\includegraphics[width=8cm]{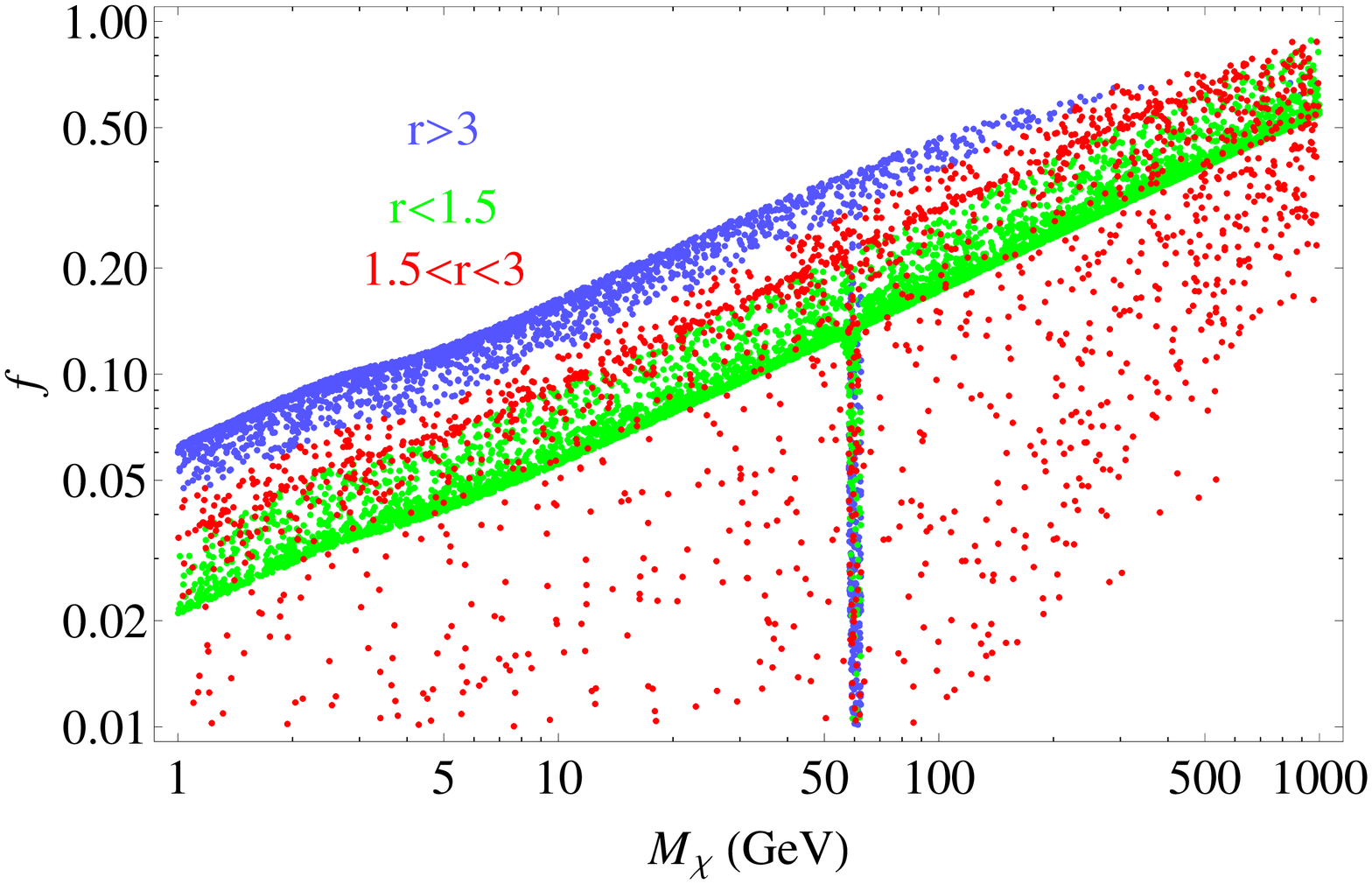}
\includegraphics[width=8cm]{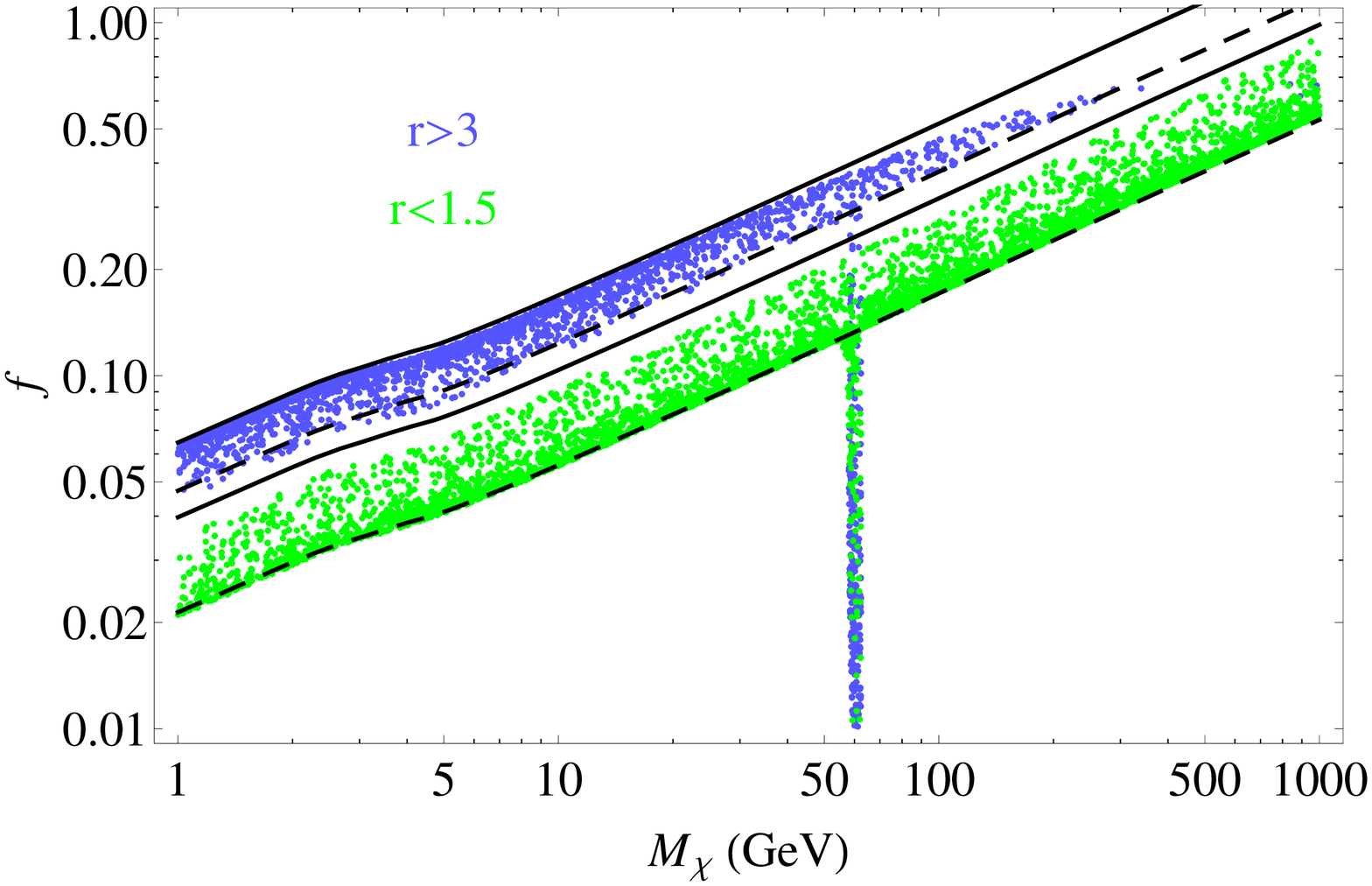}
\caption{{\it Left plot}: value of the coupling constant $f$ required to produce thermally the observed dark matter abundance for various values of the dark matter mass and the $CP$ even scalar mass, expressed as $r=m_\rho/M_\chi$. The color code denotes $r>3$, $r<1.5$ and $1.5<r<3$ for blue, green and red respectively. {\it Right plot}: The same as the left plot, but removing the points with $1.5<r<3$, to highlight the validity of the approximation  Eq.~(\ref{expf}). Besides, the solid (dashed) line shows the maximum (minimum) values of the coupling predicted by  Eq.~(\ref{expf}). }
\label{figure:fvsM}
\end{center}
\end{figure}

To check the validity of our approximations, we have calculated the values of the coupling constant $f$ versus the dark matter mass $M_\chi$ leading to the observed dark matter abundance in a scan over the four dimensional parameter space performed as described at the beginning of this Section. The result is shown in Fig.~\ref{figure:fvsM}, left panel, where we have identified with a color the value of $r$ corresponding to each point: blue, green and red for $r>3$, $r<1.5$ and $1.5<r<3$. In the right panel we have removed the points within the resonant and threshold region $1.5 < r < 3$, clearly showing the existence of two separate bands corresponding to the regimes $r<1.5$ and $r>3$. In the former case, the freeze-out is dominated by the s-wave annihilation channel into $\rho\,\eta$, whereas in the latter, by the p-wave annihilation into $\eta$ pairs. Consequently, in the case of $r > 3$, larger values of $f$ are required in order to reproduce the same relic abundance. We also show the lines corresponding to maximum (continuous) and minimum (dashed) values predicted by Eq.~(\ref{expf}) for both $r < 1.5$ and $r > 3$, assuming $x_f = 22$. As apparent from the plot, the lines  obtained using Eq.~(\ref{expf}) describe fairly well each region except for the points around $M_\chi=m_{h}/2\simeq 63$ GeV where, due to the existence of the Higgs resonance, Eq.~(\ref{expf}) does not apply.

\section{Signatures of Pseudo-Goldstone Bosons}
\label{sec:PGB}

\begin{figure}[t]
\begin{center}
\includegraphics[width=8cm]{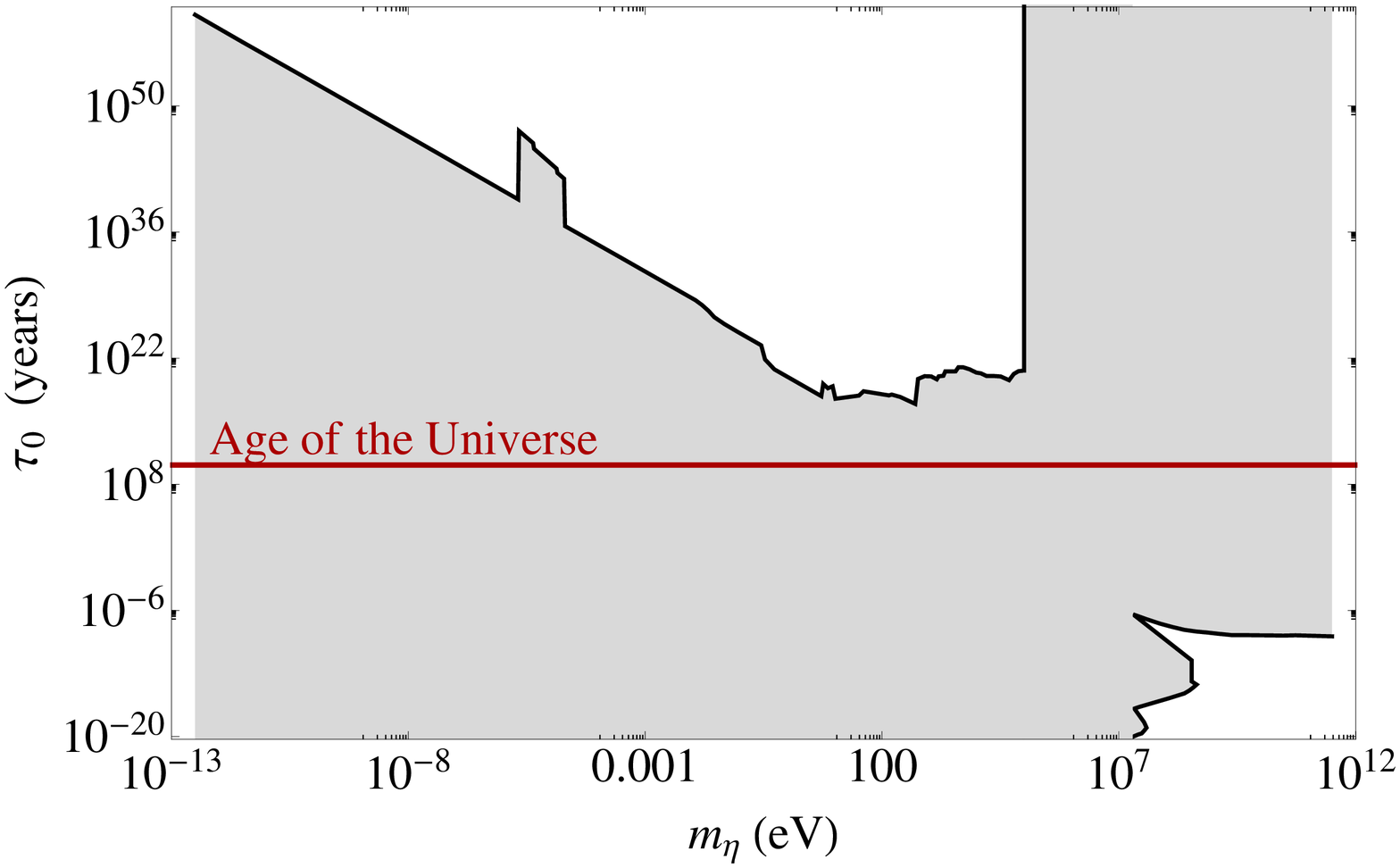}
\includegraphics[width=8cm]{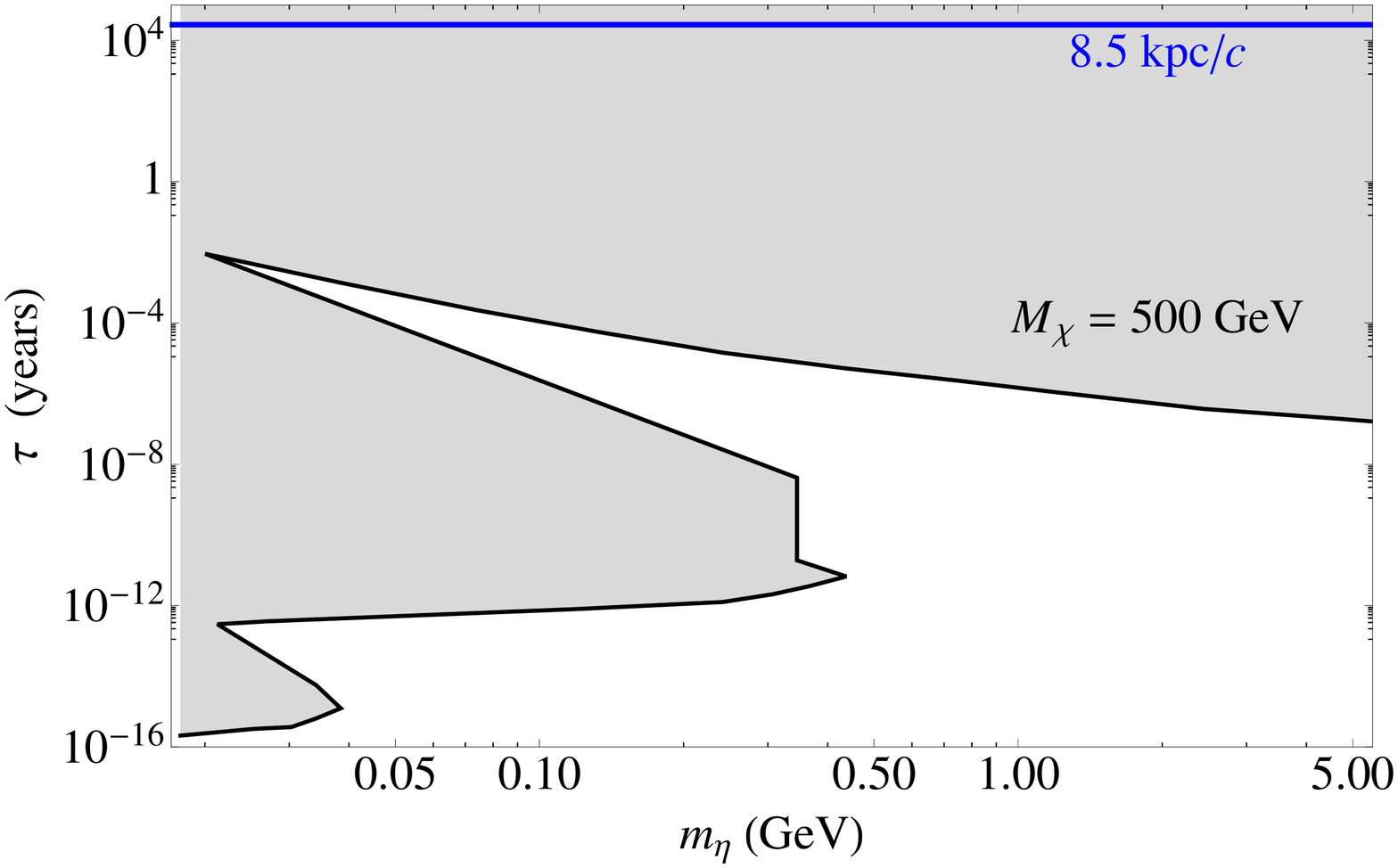}
\caption{
{\it Left plot}: allowed regions for the mass and proper lifetime of a pseudo-Goldstone boson (adapted from Fig.~6-1 of \cite{Hewett:2012ns} using Eq.~\ref{eta-width}). {\it Right plot}: allowed regions for the mass and lifetime of a pseudo-Goldstone bosons produced by dark matter annihilations assuming $M_\chi =500$ GeV, compared to the time required to reach the Earth from the Galactic center.}
\label{figure:LifeTimePlot}
\end{center}
\end{figure}

The signatures of the pseudo-Goldstone bosons in this model crucially depend on their lifetime. If the pseudo-Goldstone bosons $\eta$ are long-lived, they could have survived until the recombination era, possibly leaving their footprints in the Cosmic Microwave Background (CMB) in the form of dark radiation \cite{Weinberg:2013kea, Garcia-Cely:2013nin}. On the other hand, if they are short-lived, we could detect their decay products after being produced in dark matter annihilations, $e.g.$, in the center of our Galaxy. 

The decay rate of $\eta$ into two photons can be straightforwardly calculated from the effective Lagrangian Eq.~(\ref{Leff}), the result being:
\begin{equation}
\Gamma (\eta \to \gamma\, \gamma) = \frac{g_{\eta\gamma}^2 \,m_\eta^3  }{64 \pi } \,. 
\label{eta-width}
\end{equation}
The relevant parameters $m_\eta$ and $g_{\eta\gamma}$, or equivalently $m_\eta$ and the proper lifetime $\tau_0 ={\rm Br}(\eta\rightarrow\gamma\gamma)/\Gamma(\eta \to \gamma\, \gamma)$, are constrained by experimental searches for pseudo-Goldstone bosons. The allowed values of the pseudo-Goldstone lifetime as a function of the mass  are shown in the left panel of Fig.~\ref{figure:LifeTimePlot}, under the assumption  ${\rm Br}(\eta\rightarrow\gamma\gamma)=1$ (plot adapted from Fig.~6-1 of \cite{Hewett:2012ns}). As apparent from the plot there are two disjoint allowed regions: either the pseudo-Goldstone has a lifetime longer than $\sim 10^{20}$ years or it has a lifetime shorter than one minute. In the former case, if we assume that $\eta$ contributes to the radiation density of the Universe at the time of recombination or at Big Bang nucleosynthesis, then the pseudo-Goldstone boson must be present in the Universe also today. On the other hand, for the latter case, if the pseudo-Goldstone has a lifetime much shorter than the age of the Universe,  all the primordial pseudo-Goldstone bosons must have decayed today. Nevertheless, the model predicts a non-negligible pseudo-Goldstone production in regions with a high dark matter density, such as the Milky Way center, from the s-wave annihilations into a pseudo-Goldstone boson and a $CP$ even dark scalar, provided this annihilation channel is kinematically open. These pseudo-Goldstone bosons have an energy of the order of $M_\chi$, which implies that their lifetime (in the Galactic frame) is given by $\tau \simeq \left(M_\chi\,/m_\eta\right)\, \tau_0$. In this window $m_\eta\gtrsim 10$ MeV, hence the lifetime of the pseudo-Goldstone bosons produced in dark matter annihilations is typically much shorter than one year, as shown in Fig.~\ref{figure:LifeTimePlot}, right panel, for the particular case $M_\chi = 500$ GeV, compared to the lifetime required to reach the Earth, shown as a blue line. Therefore, pseudo-Goldstone bosons decay in flight before reaching the Earth producing a gamma-ray flux that could be  detected in gamma-ray telescopes. A similar conclusion holds for other values of the dark matter mass. In the following we analyze these two possibilities separately. 
\subsection{Long-lived pseudo-Goldstone boson scenario}
\begin{figure}[t!]
\begin{center}
\includegraphics[width=12cm]{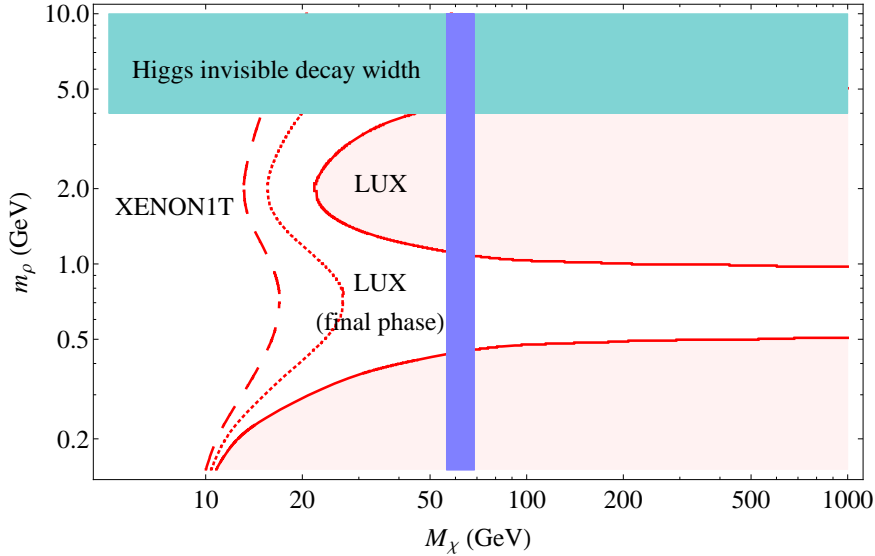}
\caption{ Excluded regions in the long-lived pseudo-Goldstone scenario from the LUX experiment and the invisible Higgs decay width, under the assumption that the dark matter particle was thermally produced and that the pseudo-Goldstone boson contributes to the effective number of neutrino species before recombination. We do not include in the analysis the Higgs resonance region (blue band).}
\label{figure:DR}
\end{center}
\end{figure}

As pointed out before, if the pseudo-Goldstone boson remains stable until the recombination era, it may contribute to the radiation energy density of the Universe, $i.e.$, it manifests itself as dark radiation.  In particular, as remarked in \cite{Weinberg:2013kea}, if it goes out of equilibrium before the annihilation of the $e^{\pm}$ pairs, but after the decoupling of most of the SM fermions, $\eta$ contributes to the relativistic number of species in the measurements of CMB \cite{Planck2013}. This effect  is quantified by the effective number of neutrino types, $N_{eff}$, present before the era of recombination \cite{Steigman}:
\begin{equation}
	N_{eff} \;=\;
	3\,+
	\,\frac{4}{7}\,\left(\frac{g_{*}\left(T_{\nu}^{d}\right)}{g_{*}\left(T_{\eta}^{d}\right)}\right)^{4/3}\,,\label{Neff}
\end{equation}
where $T^d_\nu$ and $T^d_\eta$ are the decoupling temperatures of the SM neutrinos and the pseudo-Goldstone bosons respectively, with $g_{*}\left(T_{\nu}^{d}\right)=43/4$. 

If $\eta$ decouples just before the muon annihilation epoch, then  $g_{*}\left(T_{\eta}^{d}\right)=57/4$  and consequently the effective number of neutrino species is  $N_{eff}-3=(4/7)(43/57)^{4/3}\simeq  0.39$  \cite{Weinberg:2013kea}, which is consistent within $1\sigma$ with the central value obtained in \cite{Planck2013} from combining Planck data, WMAP9 polarization data and ground-based observations of high-$\ell$, which imply $N_{\rm eff}=3.36^{+0.68}_{-0.64}$ at 95\% C.L. This scenario requires, for a given $m_\rho$, a mixing angle $\theta$ fulfilling the lower bound shown in Fig.~8 of \cite{Garcia-Cely:2013nin}, which was derived for a model with a scalar sector identical to the one under consideration. On the other hand,  under the well motivated assumption that the dark matter of our Universe was thermally produced  (which requires a coupling $f$ approximately given by Eq.~(\ref{expf})), there exists an upper limit on $\theta$ from direct search experiments and the invisible Higgs decay width, which follows from the upper limit on $f|\sin 2\theta|$ shown in Fig.~\ref{figure:fsin2thetaBound}.

Following \cite{Garcia-Cely:2013nin}, we search for allowed windows of $|\sin\theta|$ and translate them into allowed regions in the plane $m_\rho$ and $M_\chi$. The allowed regions are shown in Fig.~\ref{figure:DR}, being the pink areas excluded by the LUX experiment and the cyan area by the upper limit on  $|\theta|$  from the invisible Higgs decay width. Notice that close to the Higgs boson resonance (blue band) the limits previously derived do not apply and therefore we remove that region from our analysis.  We also report in Fig.~\ref{figure:DR} the corresponding prospects for the direct detection experiments LUX (final phase)~\cite{Akerib:2012ys} and XENON1T~\cite{Aprile:2012zx}.  For $M_\chi \gtrsim 100$ GeV, dark radiation is possible if $0.5~\GeV\lesssim m_{\rho}\lesssim 1~\GeV$. It is remarkable that a significant portion of the parameter space will be probed both by the LUX (final phase) and XENON1T experiments. For the former case, $M_\chi \gtrsim 25$ GeV might be probed, while for the latter it would be possible to probe dark matter masses as low as 15 GeV.

\subsection{Short-lived pseudo-Goldstone boson scenario}
\label{IndirectDetection}

\begin{figure}[t!]
\begin{center}
\includegraphics[width=12cm]{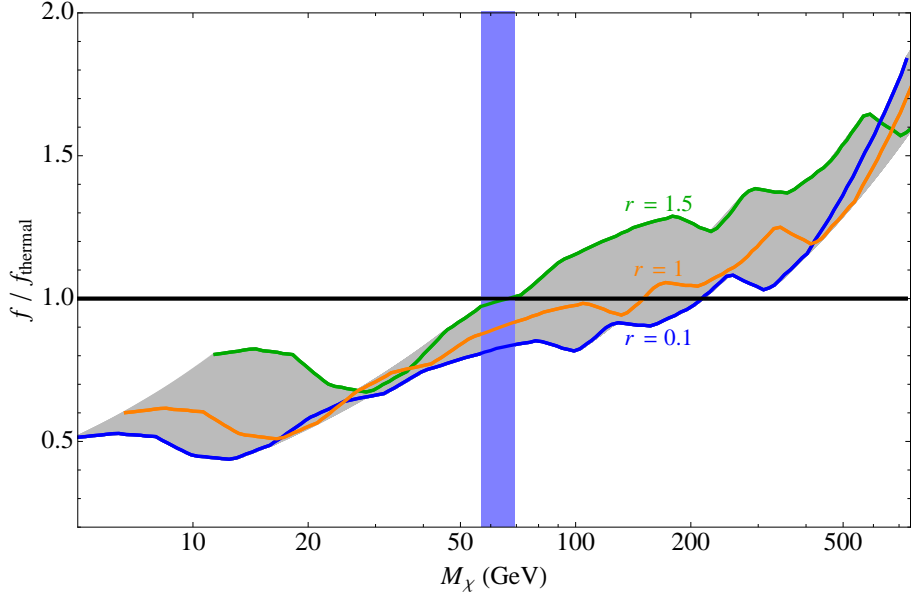}
\caption{Upper bound on the coupling constant $f$  as a function of the dark matter mass from the negative searches of gamma-ray boxes in the Fermi-LAT data, normalized to the value required to thermally produce the dark matter particles.  We assume in the plot $ \text{Br}(\eta \to \gamma\gamma) = 1$, $x_f = 20$ and vary $r$ between $0.1$ and $1.5$ (gray area). We do not include in the analysis the Higgs resonance region (blue band).} 
\label{fig:foverfth}
\end{center}
\end{figure}

This model predicts that dark matter particles might be annihilating in the center of our galaxy with a non-negligible rate, concretely through the s-wave process $\chi\chi\rightarrow \eta\rho$, if kinematically allowed. This is in contrast to the model introduced in \cite{Weinberg:2013kea}, where all the annihilation channels are p-wave suppressed \cite{Garcia-Cely:2013nin}. Therefore, this model might lead to observable signatures in indirect dark matter search experiments. Following the discussion in Section \ref{sec:production} and to allow kinematically this annihilation process we will assume in what follows that $r<1.5$.

The pseudo-Goldstone bosons produced in the annihilations $\chi\chi\rightarrow \rho\eta$ decay in flight into two photons well before reaching the Earth, as follows from Fig.~\ref{figure:LifeTimePlot}, thus generating a gamma-ray flux with a characteristic spectrum. In the center of mass frame of the annihilating dark matter particles, the energies of the $\rho$ and pseudo-Goldstone bosons are 
\begin{equation}
E_\rho = M_\chi \left(1+\frac{m_\rho^2-m_\eta^2}{4 M_\chi^2}\right)\qquad\text{and} \qquad
E_\eta = M_\chi \left(1-\frac{m_\rho^2-m_\eta^2}{4 M_\chi^2}\right)\,.
\end{equation}
whereas the energy of the photons is%
\begin{equation}
E_\gamma (\alpha) = \frac{m_\eta^2}{2 E_\eta\left(1- \cos \alpha\sqrt{1-\frac{m_\eta^2}{E_\eta^2} }  \right)}\,,
\label{Egamma}
\end{equation}
where $\alpha$ is the angle between the pseudo-Goldstone boson and the emitted photons in the annihilation frame. In the rest frame of the pseudo-scalar $\eta$ the photons are emitted isotropically, therefore the energy distribution in the galactic frame displays a characteristic box-shaped spectrum \cite{Ibarra:2012dw}, centered at $E_c \equiv (E(0)+E(\pi))/2$  and with width $\Delta E  \equiv E(0)-E(\pi)$, which are given by 
\begin{equation}
E_c = \frac{1}{2} E_\eta  \approx \frac{M_\chi}{2} \left(1-\frac{r^2}{4}\right)  \qquad\text{and}\qquad
\Delta E  = \sqrt{E_\eta^2 - m_\eta^2 } \approx M_\chi \left(1-\frac{r^2}{4}\right) \,,
\end{equation}
where it has been assumed that $m_\eta \ll m_\rho$. Namely, the center of the box is located at half the energy of the pseudo-Goldstone boson, whereas the width is given by its momentum. Besides, the dark $CP$ even scalar decays $\rho\rightarrow \eta\eta$ thus producing another contribution to the gamma-ray flux from the subsequent decay $\eta\rightarrow\gamma\gamma$. This contribution arises at lower energies, where the background is stronger, and therefore will be neglected in our analysis. The relevant part of the photon spectrum is then:
\begin{equation}
\frac{dN_\gamma}{dE_\gamma} = \frac{2}{\Delta E} \Theta \left(E_\gamma - E_c +\frac{1}{2} \Delta E \right)\Theta \left( E_c +\frac{1}{2} \Delta E -E_\gamma\right)  \text{Br} \left(\eta \to \gamma\gamma\right)
\label{spc}
\end{equation}
and the gamma-ray flux at Earth is
\begin{equation}
\phi(E_\gamma)= \frac{\langle \sigma v (\chi\chi \to \rho \eta)\rangle}{8\,\pi\,M_\chi^2}\frac{dN_\gamma}{dE_\gamma}  \frac{1}{\Delta \Omega} \int_{\Delta \Omega} d \Omega \, J_\text{ann}\,,
\label{flux}
\end{equation}
where $\Delta \Omega$ is the field of view of observation and $J_\text{ann} = \int_\text{l.o.s.} ds \, \rho_\chi^2 $ is the integral of the squared dark matter density $\rho_\chi$ along the line of sight. 

The dark matter coupling $f$ can then be constrained from searches of a box feature in the cosmic gamma-ray energy spectrum. We use the limits derived in \cite{Ibarra:2012dw} (intermediate approach), based on observations by the Fermi-LAT of the gamma ray flux from the galactic center. Those limits, derived assuming dark matter annihilation into two scalar particles of the same mass, can be appropriately adapted to our model by replacing $\langle \sigma  v \rangle \to  \langle \sigma  (\chi\chi \to \rho \eta) v \rangle (1-r^2/4)^2/2$ and $m_\text{DM}\to M_\chi(1-r^2/4)$. 

We report in Fig.~\ref{fig:foverfth} the upper bound on $f$, normalized to the values of $f$ which allow for thermal production of dark matter, given in Eq.~(\ref{expf}). In the plot we assume $ \text{Br}(\eta \to \gamma\gamma) = 1$,  $x_f = 20$ and  we vary $r$ between $0.1$ and $1.5$ (gray area). Furthermore, we highlight in blue, orange and green the bound for $r =0.1, 1 $ and $1.5$ respectively. The blue shaded area corresponds to the Higgs boson resonance, around which Eq.~(\ref{expf}) does not apply for sizable values of the mixing angle $\theta$. We then conclude that, under the assumption of thermal dark matter production and of the pseudo-Goldstone boson decaying dominantly into a pair of photons,  dark matter masses below $\sim 55\GeV$ are excluded by the Fermi data.  
\begin{figure}[t!]
\begin{center}
\includegraphics[width=13cm]{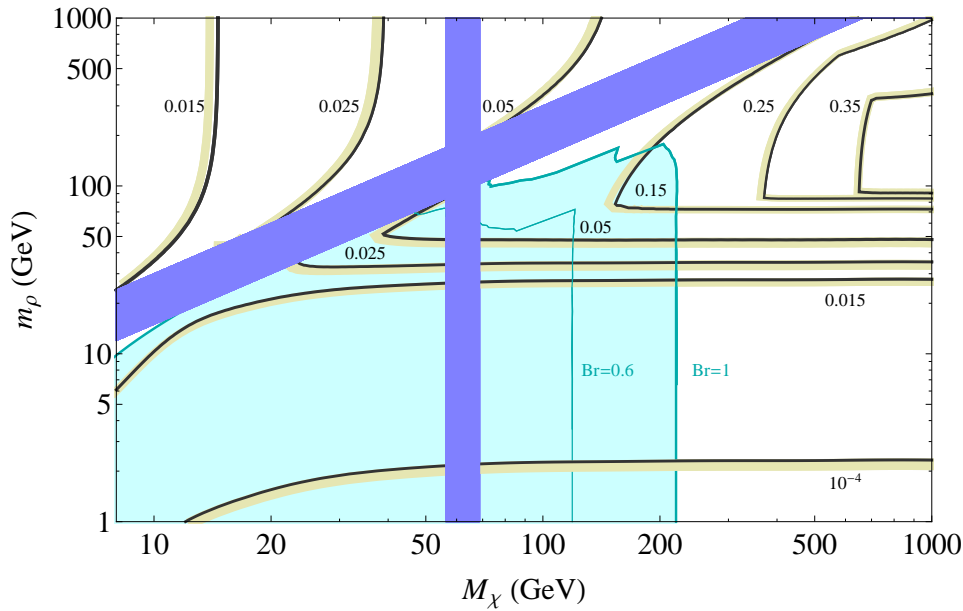}
\caption{Contours with the upper limit on the absolute value of the mixing angle $|\theta|$ in the short-lived pseudo-Goldstone scenario from the LUX experiment and the Higgs invisible decay width, under the assumption that the dark matter particle was thermally produced. The areas shaded in cyan are excluded by searches of gamma-ray boxes assuming ${\rm Br}(\eta \to \gamma\gamma)=0.6$ and $1$. We do not include in the analysis the threshold and resonance regions (dark blue bands).  }
\label{figure:ExclusionPlot}
\end{center}
\end{figure}

We summarize in Fig.~\ref{figure:ExclusionPlot} the impact of the various limits on the parameter space of this scenario under the assumption that the dark matter population in our Universe was thermally produced. Under this assumption the dark matter coupling  $f$ is determined by Eq.~(\ref{expf}) and hence the parameter space of the model is spanned by the three parameters  $M_\chi$, $m_\rho$ and $\theta$. We show in the $m_\rho$-$M_\chi$ plane the upper limit on $|\theta|$ (black lines) inferred from the LUX results and the invisible Higgs decay width (see Fig.~\ref{figure:fsin2thetaBound}). Namely, points within a given region can not have a value of the mixing angle larger than the one indicated by the corresponding label (the interior of a given region is specified by the shaded contour). Besides,  the excluded regions from gamma-ray box searches are shown in cyan for the branching ratios ${\rm Br}(\eta \to \gamma\gamma)=0.6$ and $1$.  Lastly, we remove the resonance and threshold regions, shown as dark blue bands, where the dark matter coupling $f$ cannot be univocally determined. It is remarkable that the combination of both direct and indirect detection experiments can probe, and possibly exclude, a large portion of the parameter space of the model for light $CP$ even dark scalars, concretely when $m_\rho\lesssim 1.5 M_\chi$. On the other hand, when $m_\rho\gtrsim 3 M_\chi$ the kinematically accessible dark matter annihilation channels are all p-wave suppressed leading to no observable signature in indirect dark matter searches. In this region of the parameter space, however, signals could be detected in direct dark matter searches or in the invisible Higgs decay width.

\section{Conclusions}
\label{sec:conclusions}

We have presented a model where the Standard Model is extended by a dark sector consisting in a chiral fermion and a complex scalar with charges 1 and 2, respectively, under a global continuous $U(1)$ symmetry. The global symmetry is assumed to be spontaneously broken by the vacuum expectation value of the complex scalar, thus leading to a Goldstone boson (or a massive pseudo-Goldstone boson if the symmetry is not exact) which is a candidate of dark radiation. Furthermore, the symmetry breaking leads to a residual $Z_2$ symmetry under which the chiral fermion is odd. Hence, the chiral fermion is absolutely stable and a candidate of dark matter. The model predicts possibly sizable contributions to the invisible Higgs decay width and to the scattering rate of dark matter particles off nuclei, thus allowing to constrain the model parameters with experiments.

We have analyzed the thermal production of dark matter particles in the early Universe and we have found that the annihilation into a $CP$ even dark scalar and a (pseudo-)Goldstone boson proceeds in the s-wave, while all other annihilation processes are p-wave suppressed. Therefore, when kinematically open, this annihilation channel determines the dark matter relic abundance. On the other hand, when kinematically closed, the relic abundance is determined by the p-wave annihilation into two (pseudo-)Goldstone bosons. For each of the cases we have found approximate expressions for the coupling constant that leads to the observed dark matter abundance today. The existence of a s-wave annihilation channel is due to the explicit $C$ and $P$ breaking induced by the chiral fermion and does not arise in models where the dark matter particle is a Dirac fermion. As a result, the phenomenology of the model with chiral fermions as dark matter particles is qualitatively different to the one with Dirac fermions.

We have then focused on the case in which the global $U(1)$ symmetry is not exact, hence the Goldstone boson is massive and decays into two photons. The mass and lifetime of the pseudo-Goldstone boson are constrained by various experiments. There are at present two allowed windows, one with a lifetime longer than $\sim 10^{20}$ years and one with a lifetime shorter than one minute. We have analyzed the experimental signatures of the pseudo-Goldstone bosons in those two windows and analyzed the interplay with the limits from thermal production, the invisible Higgs decay width and direct dark matter searches. In the former scenario, the pseudo-Goldstone boson is a candidate of dark radiation. If this is the case, and assuming that the dark matter particle was thermally produced, the direct search experiments LUX and XENON1T could find a positive signal if the dark matter mass is larger than 25 GeV or 15 GeV, respectively. In the latter scenario, on the other hand, \hspace{25pt}s-wave dark matter annihilations in the Galactic center into a pseudo-Goldstone boson and a $CP$ even dark scalar produce, if kinematically allowed, an intense gamma-ray flux displaying a box shaped spectrum. We have determined the limits on this scenario from the Fermi-LAT data and we have found that, if the $CP$ even scalar is much lighter than the dark matter, gamma-ray measurements exclude dark matter masses below 220 GeV ( 120 GeV) when ${\rm Br}(\eta\rightarrow \gamma\gamma)=1$ (0.6).

\section*{Acknowledgements}
We are grateful to Miguel Pato and Hyun Min Lee for useful discussions. 
This work was supported in part by the DFG cluster of excellence ``Origin and Structure of the Universe'', by the ERC Advanced Grant project ``FLAVOUR''(267104) (A.I., E.M.) and by the Graduiertenkolleg ``Particle Physics at the Energy Frontier of New Phenomena'' (C.G.C.).

\section*{Appendix: Cross-sections}

In this appendix we report the annihilation cross-sections of $\chi$ for an arbitrary center of mass energy $\sqrt{s}$ in the limit $\theta=0$.
We introduce for convenience the notation

\begin{eqnarray}
t = \frac{\sqrt{s}}{M_\chi}\,,\quad\quad\quad
r = \frac{m_{\rho}}{M_{\chi}}\,,\quad\quad\quad \gamma_{\rho}= \frac{\Gamma_{\rho}}{M_{\chi}}\,\nonumber
\end{eqnarray}
and we define the functions
\begin{eqnarray}
        K_{\eta\eta}(t)  =  \sqrt{-4+t^{2}}\,,&&
        K_{\rho\eta}(r,t)  =  \sqrt{\frac{\left(r^{2}-t^{2}\right)^{2}\left(-4+t^{2}\right)}{t^{2}}}\,,\nonumber\\
        K_{\rho\rho}(r,t)  &=&  \sqrt{\left(-4\,r^{2}+t^{2}\right)\left(-4+t^{2}\right)}\,.\nonumber
\end{eqnarray}
In terms of these definitions, the annihilation  cross-sections are the following:
\begin{eqnarray}
&&\sigma\left(\chi\,\chi\to \eta\,\eta\right) \; = \;
 \frac{f^4 \left(K_{\eta \eta }(t) \left(r^4 t^2-4 r^2 \gamma _{\rho }^2-4 t^4\right)-2 \,t  \left(-r^4+r^2
   \gamma _{\rho }^2+t^4\right)\right)\log \left(\frac{t-K_{\eta \eta }(t)}{t+K_{\eta \eta }(t)}\right)}
   {64\, \pi\, M_{\chi }^2\, t\, \left(t^2-4\right)  \left(r^2\, \gamma _{\rho }^2+\left(t^2-r^2\right)^2\right)}\,,\nonumber\\
&&\sigma\left(\chi\,\chi\to \rho\,\eta\right) \; = \;
\frac{f^4 \left(r^4 K_{\rho \eta }(r,t)+2 \left(2 r^4-3 r^2 t^2+t^4\right) \log \left(\frac{r^2-t^2-K_{\rho \eta }(r,t)}{r^2-t^2+K_{\rho \eta }(r,t)}\right)\right)}
{32 \,\pi\,  M_{\chi }^2\,   t^4 \left(t^2-4\right)}\nonumber\,,
\end{eqnarray}

\begin{eqnarray}
&&\sigma\left(\chi\,\chi\to \rho\,\rho\right) = 
 \frac{f^4 }{128\,  \pi \,M_{\chi }^2\, t^2 \left(t^2-4\right) }\nonumber\\
&&\left(\frac{2\,K_{\rho \rho }(r,t)\, \left(9 r^8 \left(t^2-2\right)+r^6 \left(80-48 t^2\right)+r^4 \left(3 t^4+16 t^2-32\right)+16 r^2 t^2 \left(t^2+4\right)-4 t^4
   \left(t^2+8\right)\right) }{\left(r^2-t^2\right)^2 \left(r^4-4 r^2+t^2\right)}\right.\nonumber\\
&&-\left.\frac{4 \left(18 r^6+10 r^4 \left(t^2-8\right)+r^2 \left(-11 t^4+16
   t^2+32\right)+t^2 \left(t^4+16 t^2-32\right)\right) \log \left(\frac{t^2-2 r^2-K_{\rho \rho }(r,t)}{t^2-2 r^2+K_{\rho \rho }(r,t)}\right)}{2 r^4-3 r^2 t^2+t^4}\right)\,.
   \nonumber
\end{eqnarray}

\end{document}